\documentclass[final,5p,times,twocolumn]{elsarticle}

\usepackage{graphicx}
\usepackage{amssymb}
\usepackage{amsmath}
\usepackage{color}
\usepackage{lineno}
\usepackage{bigstrut}
\usepackage{multirow}
\usepackage{setspace}

\usepackage{lineno}

 \journal{Journal Name}

\begin{document}

\begin{frontmatter}


\title{Grazing incidence-X-ray fluorescence for a dimensional and elemental characterization of well-ordered nanostructures}



\author[label1]{Philipp H\"onicke}  
\author[label1]{Anna Andrle}
\author[label1]{Yves Kayser}
\author[label1]{Konstantin V.~Nikolaev}
\author[label2]{J\"urgen Probst}
\author[label1]{Frank Scholze}
\author[label1]{Victor Soltwisch}
\author[label3]{Thomas Weimann}
\author[label1]{Burkhard Beckhoff}

\address[label1]{Physikalisch-Technische Bundesanstalt, Abbestr. 2-12, 10587 Berlin, Germany}
\address[label2]{Helmholtz-Zentrum Berlin (HZB), Albert-Einstein-Str. 15, 12489 Berlin, Germany}
\address[label3]{Physikalisch-Technische Bundesanstalt, Bundesallee 100, 38116 Braunschweig, Germany}

\begin{abstract}
The increasing importance of well-controlled ordered nanostructures on surfaces represents a challenge for existing metrology techniques. To develop such nanostructures and monitor complex processing constraints fabrication, both a dimensional reconstruction of nanostructures and a characterization (ideally a quantitative characterization) of their composition is required. 
In this work, we present a soft X-ray fluorescence-based methodology that allows both of these requirements to be addressed at the same time. By applying the grazing-incidence X-ray fluorescence technique and thus utilizing the X-ray standing wave field effect, nanostructures can be investigated with a high sensitivity with respect to their dimensional and compositional characteristics. By varying the incident angles of the exciting radiation, element-sensitive fluorescence radiation is emitted from different regions inside the nanoobjects. By applying an adequate modeling scheme, these datasets can be used to determine the nanostructure characteristics.
We demonstrate these capabilities by performing an element-sensitive reconstruction of a lamellar grating made of Si$_3$N$_4$, where GIXRF data for the O-K$\alpha$ and N-K$\alpha$ fluorescence emission allows a thin oxide layer to be reconstructed on the surface of the grating structure. In addition, we employ the technique also to three dimensional nanostructures and derive both dimensional and compositional parameters in a quantitative manner.
\end{abstract}


\end{frontmatter}


\section{Introduction}
\label{S:1}
Well-controlled and well-defined nanostructures and their achievable properties enable many modern applications in science and industry \cite{Pala2013,M.L.Brongersma2014,Zheng_2016}. The most prominent example is probably the ongoing pursuit of Moore's law\cite{Moore1965} in the semiconductor industry. Here, the technological advances in the fields of lithography \cite{Wagner_2010} and related nanotechnologies results in an explosion of complexity of modern semiconductor device structures. This increase in complexity is due to decreasing critical feature dimensions\cite{S.W.King2013,S.Natarajan2014,Kim_2018}, incorporation of many different materials\cite{Clark2014} and the growing importance of the third dimension for a further densification of the structures\cite{Vandooren_2018,Ryckaert_2018}. Similar trends are also relevant for related applications of nanostructures. 
Manufacturing such nanostructures requires metrologically rigorous methods in order to ensure reasonable yields \cite{S.W.King2013,Bunday2016,Bunday2017}. The performance of such complex 2D and 3D devices is highly dependent on the precision of the dimensional parameters and on the chemical composition and spatial element distributions on a nanometric scale, e.g. for dopant depth profiles and barrier layer thicknesses. 3D metrology plays a crucial role in these developments and in manufacturing process control. 
Typical metrology techniques that can be added to the existing "toolset" in this context include microscopy-based techniques for scanning electron microscopy (SEM) and transmission electron microscopy (TEM), on the one hand, and secondary ion mass spectroscopy (SIMS)\cite{Franquet_2016, W.Vandervorst2016} and atom probe tomography (APT)\cite{W.Vandervorst2016} on the other. The latter two techniques can resolve 3D nanostructures with sub-20 nm spatial resolution as well as their elemental composition. However, especially APT requires a tedious sample preparation and both techniques are destructive. In addition, it can be rather time consuming to derive statistically relevant information by analyzing numerous nominally identical objects.

Also X-ray based techniques, especially Critical Dimension Small Angle X-Ray Scattering (CDSAXS) \cite{Sunday_2015,Kline_2017}, are being applied to the field of nanostructure metrology. CDSAXS is a transmission scattering measurement where the sample is rotated to probe the vertical profile, allowing 2D and 3D reconstructions of the dimensions and composition of periodic nanostructures to be performed. However, this technique requires the sample to be thin enough for obtaining a reasonable transmission. A slightly different approach is utilized in GISAXS (grazing-incidence SAXS), which is currently being evaluated as a metrology technique for lithographically patterned features such as line gratings \cite{Soltwisch2017}. However, both in GISAXS and CDSAXS, there is typically no optical contrast between different materials present in the nanostructure and thus only dimensional parameters can be derived.

Another technique that has been shown to add capabilities for probing 2D and 3D nanostructures with respect to both their dimensional parameters and their elemental composition is grazing-incidence X-ray fluorescence (GIXRF) analysis \cite{M.Dialameh2017,V.Soltwisch2018}. This non-destructive technique is based on a variation of the incident angles between the nanostructured sample surface and the incoming monochromatic X-ray beam. These angular variations induce changes within the X-ray standing wave field (XSW) intensity distribution in all three dimensions and occur due to interference between incident and reflected radiation. These local intensity modulations serve as the nanoscale sensor for both the dimensional properties of the nanoobject and its elemental composition due to the elemental specificity of the X-ray fluorescence technique. 

In this work, we have applied the reference-free GIXRF methodology of PTB \cite{Hoenicke2019,M.Mueller2014}, which provides a {SI} traceable quantitative access to the amounts of material within the nanostructures, to different artifical two- and three-dimensional nanostructures. Using silicon nitride grating nanostructures, we demonstrate the elemental sensitivity of the methodology as we can distinguish the fluorescence signals from nitrogen and oxygen atoms. The atoms are differently distributed within the nanostructure studied, and thus a very different angular behavior of their fluorescence emission is observed allowing an element-sensitive reconstruction of the structure. In addition, we apply the technique to three-dimensional chromium nanostructures to demonstrate the applicability for more complex and technologically relevant nanostructures. 

\section{Experimental}
\label{S:2}
\begin{figure}[h]
\centering
\includegraphics[width=8cm]{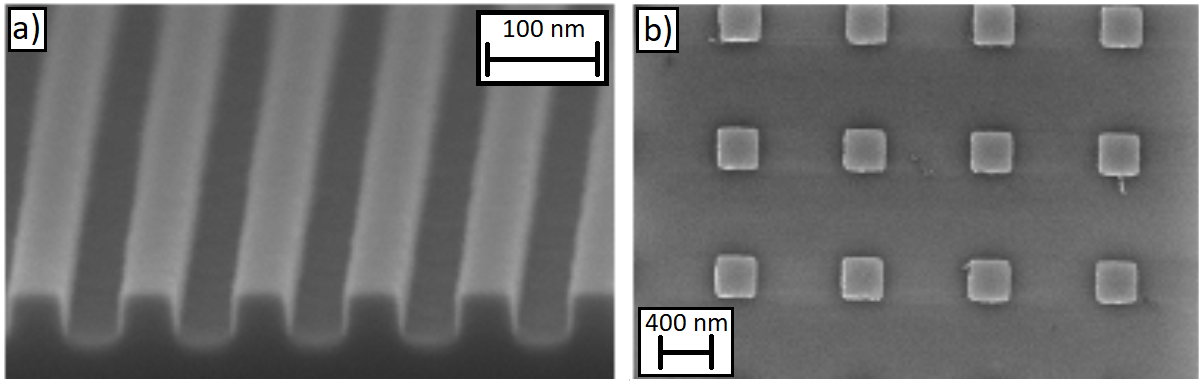}
\caption{Scanning electron microscopy images of the different nanostructured samples used in this work; part a) shows the Si$_3$N$_4$ lamellar grating and part b) shows the 3D chromium structures.}
\label{fig:SEMs}
\end{figure}

As an example of two-dimensional nanostructures, we use two different lithographically structured silicon nitride grating layers on a silicon substrate. Both Si$_3$N$_4$ lamellar gratings were manufactured by means of electron beam lithography (EBL) at the Helmholtz-Zentrum Berlin. Both gratings have a nominal pitch of 100 nm and line height of 90 nm; one has a line width of 40 nm (sample A) while the other has line width of 50 nm (sample B). The grated areas on both samples are 1 mm by 15 mm in size, with the grating lines oriented parallel to the long edge. The left side of figure \ref{fig:SEMs} shows an SEM cross-section image of a witness grating structure. In the surrounding areas, the original Si$_3$N$_4$ layer is left untouched. To manufacture the gratings, a silicon substrate with a chemical-vapor deposited 90 nm-thick Si$_3$N$_4$ layer was patterned using EBL (including oxygen plasma resist stripping). Further details about the fabrication procedure can be found in ref. \cite{V.Soltwisch2018}.  

As an example of three-dimensional nanostructures, we use an array of chromium cuboids with a nominal dimension of 300 nm x 300 nm x 20 nm (width x length x height) on a 300 nm SiO$_2$ on Si substrate, see figure \ref{fig:SEMs}. The well-ordered Cr structures were fabricated using the EBL technique on an area of 1 mm by 15 mm. Further details about the fabrication procedure can be found in ref. \cite{M.Dialameh2017}.

The reference-free GIXRF measurements on the Si$_3$N$_4$ gratings were carried out on the plane-grating monochromator (PGM) beamline \cite{F.Senf1998} for undulator radiation in the PTB laboratory \cite{B.Beckhoff2009c} at the BESSY II electron storage ring. The PGM beamline provides soft X-ray radiation in the photon energy range of 78 eV to 1860 eV with high spectral purity. The Cr nanostructures were measured on the four crystal monochromator (FCM) \cite{Krumrey1998} beamline. This beamline provides monochromatized X-ray radiation in the photon energy range from around 1.7 keV up to 10 keV. On both beamlines, the GIXRF experiments were conducted employing the radiometrically calibrated instrumentation of PTB allowing a reference-free quantification \cite{Beckhoff2008} of elemental mass depositions present on the sample. An ultrahigh-vacuum (UHV) chamber equipped with a 9-axis manipulator \cite{J.Lubeck2013} was used for the measurements, allowing for variations of both the incident angle $\theta$ (between the X-ray beam and the sample surface) and the azimuthal incidence angle $\varphi$ (defined as the angle between the lines of the grating structure and the plane of incidence) with respect to the grating lines. The incident angle $\theta$ can be aligned with an uncertainty well below $0.01^{\circ}$, which is sufficient for the GIXRF experiments. The azimuthal incidence angle $\Phi_i$ can also be aligned with similar accuracy using X-ray reflectometry (XRR) with photodiodes mounted on a 2$\theta$ axis.

During the GIXRF experiments, the fluorescence radiation emitted from the sample is detected by means of a calibrated silicon drift detector (SDD)\cite{F.Scholze2009}, which is placed in the orbital plane of the electron storage ring and perpendicular to the incident X-ray beam direction in order to minimize the radiation scattered from the sample. The incident photon flux is monitored by means of a calibrated photodiode. 

The reference-free GIXRF experiments on the Si$_3$N$_4$ lamellar grating were carried out at an incident photon energy of 680 eV (in contrast to our previous work \cite{V.Soltwisch2018}). This photon energy allows O-K${\alpha}$ fluorescence radiation to be excited in addition to the N-K${\alpha}$ fluorescence. Using this additional signal, the sensitivity of the reconstruction with respect to the surface oxide layer (to be expected due to the oxygen plasma treatment after etching) can be improved, as this layer now provides a direct signal.  

\begin{figure}[h]
\centering
\includegraphics[width = 0.9\linewidth]{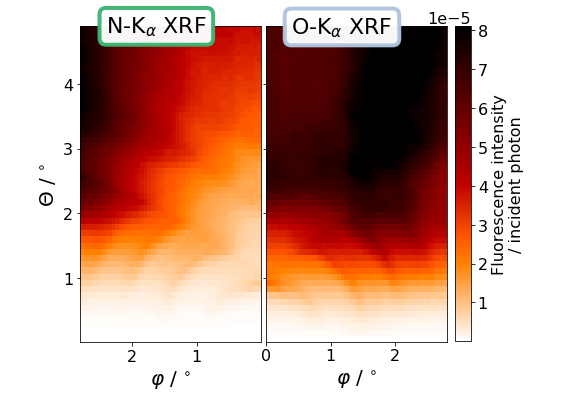}
\caption{Comparison of the experimental N-K fluorescence intensities (a) with the O-K fluorescence intensities (b) as a function of the incident angle $\theta$ and the azimuth angle $\varphi$ for the sample with a nominal line width of 40 nm (sample A).}
\label{fig:map_N_O_comp}
\end{figure}   

\begin{figure}[h]
\centering
\includegraphics[width=7cm]{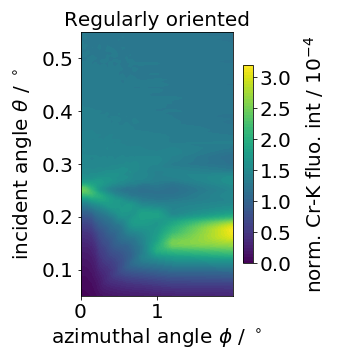}
\caption{Cr-K fluorescence intensities as a function of the incident angle $\theta$ and the azimuth angle $\varphi$ for the 3D Cr nanostructures.}
\label{fig:Cr_comparison}
\end{figure}

The 3D Cr nanostructure was excited using a photon energy of 7 keV and both the incident and the azimuthal angles were scanned. The angle-dependent Cr-K fluorescence intensities (both the K${\alpha}$ and K${\beta}$) derived for the Cr sample are shown in figure \ref{fig:Cr_comparison}. The 3D nanostructures show a strong dependence on the fluorescence signal with respect to the azimuthal angle $\varphi$. 

For both sample types, the angle of incidence $\theta$ was scanned at different positions of the azimuthal angle $\varphi$. At each position, an X-ray fluorescence spectrum was recorded using a windowless SDD. The spectra were deconvoluted using the known detector response functions \cite{F.Scholze2009} for the relevant fluorescence lines of nitrogen, oxygen or chromium and relevant background contributions. The resulting count rates $F(E_0,\theta,\varphi)$ for the fluorescence line of interest are normalized with respect to the sine of the incident angle $\theta$, the incident photon flux $N_0$, the effective solid angle of detection $\frac{\Omega}{4\pi}$ and the detection efficiency $\epsilon_{E_f}$ of the SDD for the respective fluorescence photons in order to derive the emitted fluorescence intensity $\Phi_i (E_0,\theta,\varphi)$ in accordance with equation \ref{eq:1}. 

\begin{equation}
\Phi_i (E_0,\theta,\varphi)= \frac{4\pi\sin\theta}{\Omega}\frac{F(E_0,\theta,\varphi)}{N_0\epsilon_{E_f}}  
\label{eq:1}
\end{equation}

\section{Results and Discussion}
\subsection{Characterization of the 2D gratings}
The angle-dependent experimental datasets for nitrogen and oxygen can be used for a quantitative reconstruction of the dimensional parameters and the elemental distributions within the nanostructure. As shown in figure \ref{fig:map_N_O_comp}, the two fluorescence signal intensities from sample A differ in their behavior, particularly in the higher angle regime. This originates from the fact that the oxygen atoms present in the nanostructure are distributed differently (mainly at the surface) than the nitrogen atoms providing additional experimental input to the modeling. The experimental data can be modeled by employing the 2D Sherman equation \cite{V.Soltwisch2018} (see eq. \ref{eq:2}) and a calculation of the angle-dependent intensity distributions $I_{XSW}$ of the X-ray standing wave using a finite element-based solution of Maxwell's equations 
Here, we use \textit{JCMWave}\cite{burger_jcmsuite:_2008} for this procedure. Then, we numerically integrate:
\begin{equation}
\Phi_i (E_0,\theta,\varphi) = \frac{W_i \rho \tau_{E_0} \omega_k}{D_x} 
\sum_{ij}
I_{XSW}(x_i,y_j) e^{-\rho\mu_{E_i}y_e(x_i,y_j)} \,\delta x\,\delta y
\label{eq:2}
\end{equation}
where $D_x$ is the lateral grating period (pitch) and $y_e$ is the distance from each point to the surface of the structure (towards the SDD), which is needed for self-attenuation correction. The finite-element algorithm can vary the line shape profile in order to reproduce the experimental data. Additional constant experimental parameters such as beam divergence and material-specific parameters such as optical constants, the weight fraction $W_i$ of the relevant element $i$ (here the fraction of nitrogen in Si$_3$N$_4$) and the fluorescence production cross section $\tau_{E_0} \omega_k$ (taken from databases \cite{T.Schoonjans2011} and dedicated experiments \cite{P.Hoenicke2016a}) at the excitation photon energy $E_0$ and shell $k$ are also employed. We use a model in which the line height and width, the surface oxide layer thickness on the grating line, the surface oxide layer thickness in the groove and the sidewall angle are parameterized to allow the grating line profile to be easily changed. The Si$_3$N$_4$ grating line width is defined at the half-height of the finite element model. The thickness of residual Si$_3$N$_4$ in the grooves can also be implemented in the model but this parameter was negligible here. In addition, a native SiO$_2$ layer, which is to be expected between the Si wafer and the Si$_3$N$_4$ was also added but its thickness was not changed during modeling. Further details on the model can be found in ref.\cite{Andrle_2019}. Based on this model, the FEM solver calculates the electric field strength at each position inside the structure, which, by taking into account self-attenuation on the way out ($e^{-\rho\mu_{E_i}y_e(x_i,y_j)}$) is numerically integrated in order to gain absolutely emitted fluorescence intensities. These are then compared against the experimental data.

As already mentioned in our previous work, such an FEM based calculation of the XSW field intensities requires significant computational resources \cite{Andrle_2019}. One possible way to reduce the expected computational resources is to reduce the degrees of freedom of the model or to at least reduce the size of the parameter space to be searched. We do this by keeping each material's optical constants fixed instead of varying them in order to take non-bulk material properties into account. The optical constants in use were determined in XRR experiments on the non-structured sample area next to the grating structure.

\begin{figure}[h]
\centering
\includegraphics[width = 1\linewidth]{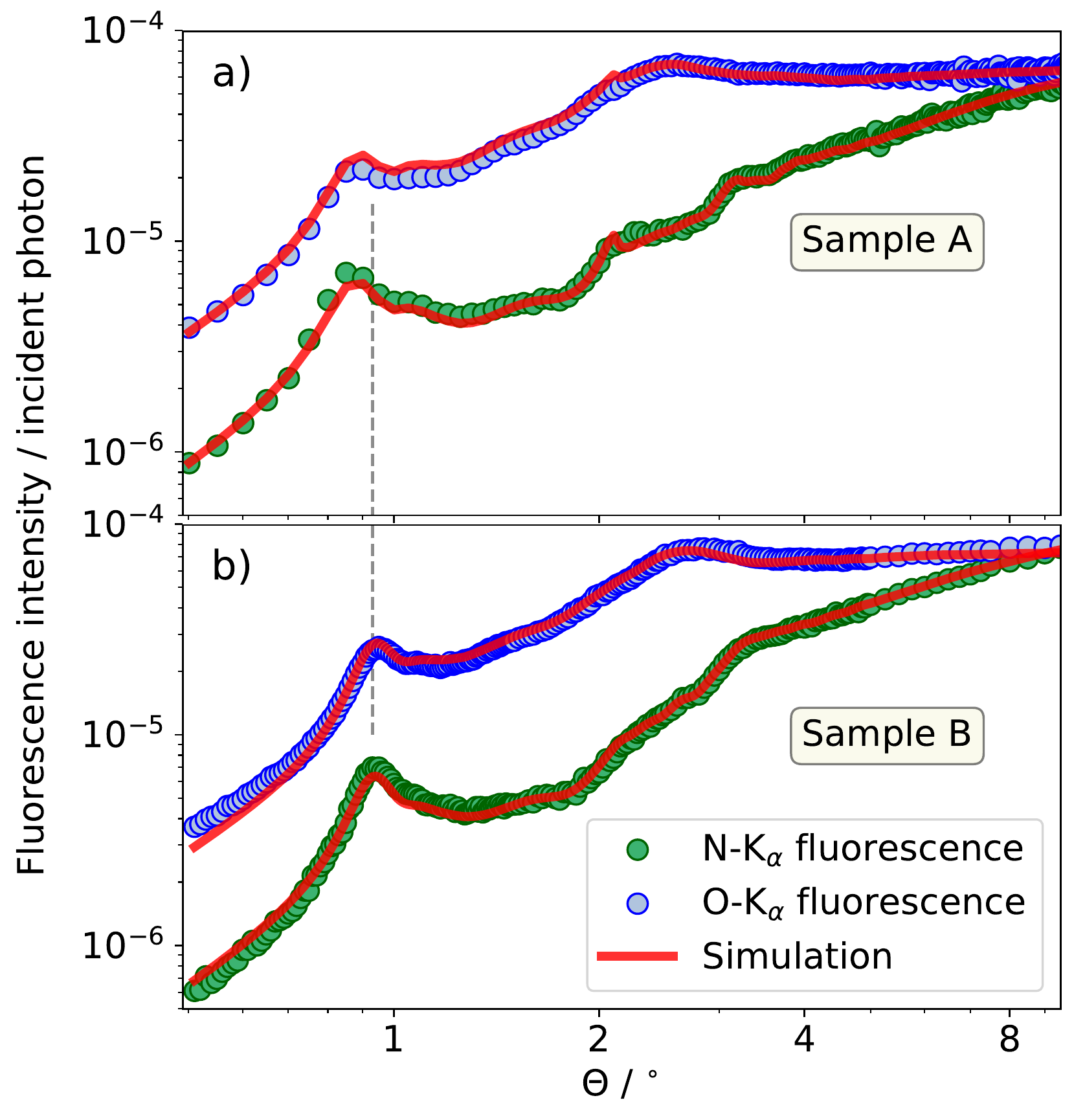}
\caption{Comparison of the normalized N- and O-datasets (dotted lines) for both samples and the corresponding calculation results (red solid lines) at an azimuthal angle of $\varphi = 0^{\circ}$. The dashed vertical line depicts the changing position of the first maximum and thus the sensitivity on the nanostructure geometry.}
\label{fig:Comp_N_vs_O}
\end{figure}

In figure \ref{fig:Comp_N_vs_O}, comparisons of the normalized N- and O-datasets (circles) for both samples and the corresponding best-fit calculation results (red solid lines) at an azimuthal angle of $\varphi = 0^{\circ}$ are shown. A comparison of the angular fluorescence profiles for N-K${\alpha}$ and O-K${\alpha}$ shows significant differences, which are due to the different distribution of these two elements within the nanostructure. Only if the two relevant materials (SiO$_2$ and Si$_3$N$_4$) are distributed correctly (both with respect to their spatial position in the structure and with respect to their overall amount) in the field simulation and, consequently, the correct regimes of the XSW are integrated, can these two datasets be modeled. The prominent feature at an incident angle of about $\theta = 1^{\circ}$ demonstrates the sensitivity on the dimensionality of the nanostructure; for both fluorescence signals, the peak position changes with increasing line width between samples A and B (see figure \ref{fig:Comp_N_vs_O}).

The dimensional parameters used for the calculation of the red solid lines in figure \ref{fig:Comp_N_vs_O} are as follows: for both samples, a sidewall angle of 82$^{\circ}$ was found. The linewidths are 44 nm (sample A) and 51 nm (sample B), whereas the heights are about 90 nm (sample A) and 100 nm (sample B). The surface oxide layer on the Si$_3$N$_4$ grating was found to be 2.7 nm (sample A) and 3.5 nm (sample B) whereas the groove oxide layer thickness is 2.5 nm (sample A) and 5.3 nm (sample B). Even though the overall agreement between the calculated and the experimental data is already good and all main trends are reproduced, the agreement is not satisfactory for some features of the experimental data. As shown in figure \ref{fig:Comp_N_vs_O}, the intensities calculated for the first peak of about $\theta = 0.9^{\circ}$ are too low for the nitrogen data; furthermore, the peak at $\theta = 2.5^{\circ}$ for the oxygen data is still inaccurate in the calculation. This indicates that the model used is not yet in perfect agreement with the real structure. However, this also indicates that the discrimination capability of the technique is high, as a model which is only close to reality cannot fully describe the experimental dataset. There are two likely reasons for the remaining discrepancies. First, until this point, the oxide layer has only been implemented as a box-like layer with a sharp interface towards the Si$_3$N$_4$.  It is more likely that there is a smooth transition between the two materials, which will be implemented in future research. Second, one drawback of the GIXRF technique is that each angle combination for $\theta$ and $\varphi$ is measured separately and thus the sample is irradiated for a long time. This can result in the growth of an additional carbonaceous surface contamination layer, which can be accounted for by including the fluorescence emission from carbon in future research.

\subsection{Characterization of the 3D nanostructures}
The well-ordered 3D Cr nanostructures used in this work can, in principle, be modeled by applying the FEM based approach. However, for practical reasons, the FEM based calculation of the XSW cannot be easily applied here. Due to the dimensional parameters of the Cr nanostructures, the computational volume of the FEM needs to be larger than 1000 nm x 1000 nm x 20 nm. At the finite element size required, this results in a very large number of finite elements; thus, the calculation consumes far too many resources for it to be performed at reasonable speeds, even on large computers. As an alternative approach, we apply a many beam dynamical diffraction theory (MB-DDT) \cite{Nikulaev2019} approach.

\begin{figure}[h]
\centering
\includegraphics[width=8cm]{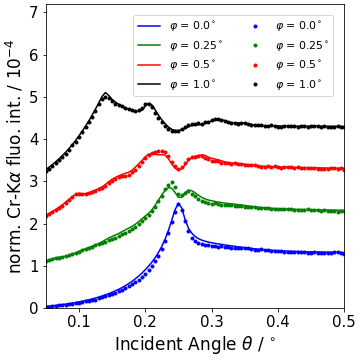}
\caption{Comparison of the normalized Cr-K$\alpha$ fluorescence intensities as a function of the incident angle $\theta$ for various azimuth angles as well as best fit calculations (see text).}
\label{fig:Cr_fitcomparison}
\end{figure}   

In addition, the Sherman equation is reformulated as follows for 3D nanostructures:
\begin{equation}
    \Phi_i (E_0,\theta,\varphi) = \frac{W_i \rho \tau_{E_0} \omega_k}{D_x D_y} 
    \iiint\displaylimits_V
    I_{XSW}(x,y,z) e^{-\rho\mu_{E_i}z_e(x,y,z)} \,dV
    \label{eq:4}
\end{equation}
The numerical integration is performed for each volume element $dV$ and a correction of the self-attenuation of the fluorescence signal is taken into account. This correction can be performed similarly to the 2D case, with $z_e(x,y,z)$ being the distance to the structure surface in the direction of the SDD. However, for the structures employed here, the self-attenuation correction can be omitted, as the high-energy Cr fluorescence photons are practically not attenuated. In \cite{Nikulaev2019}, this integral is computed analytically using the semi-analytic solution of Maxwell's equations. This allows computational resources to be drastically reduced compared to the FEM approach.

\begin{figure}[!h]
\centering
\includegraphics[width=9cm]{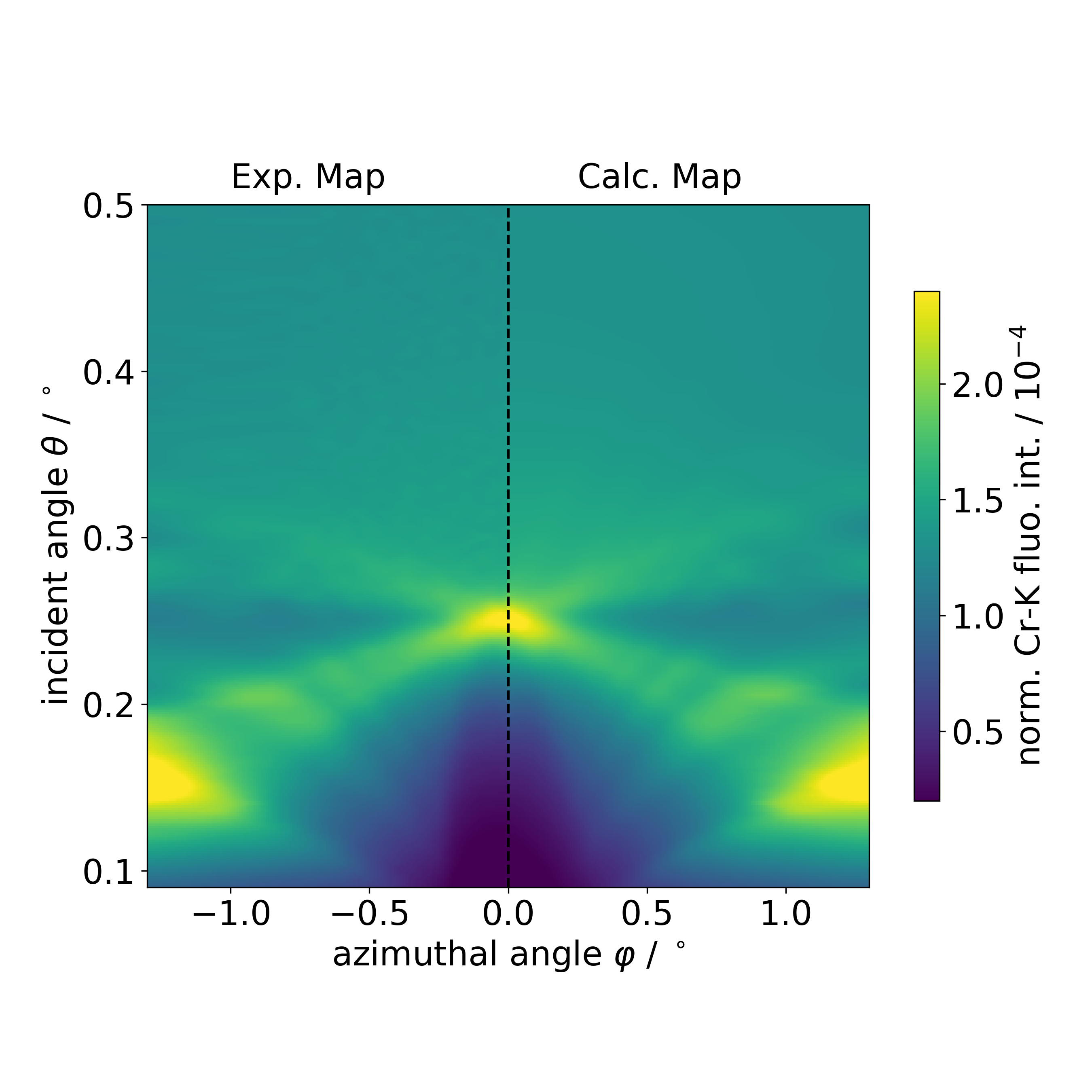}
\caption{Comparison of the experimental Cr-K$\alpha$ fluorescence intensity map (left side) as a function of the azimuthal angle $\varphi$ and the incident angle $\theta$ and a calculated map for the best fit parameters using the MB-DDT approach (right side).}
\label{fig:Cr_fitcomparison_Maps}
\end{figure}

The Cr nanostructures can then be modeled by optimizing the dimensions of the Cr boxes (width and height) and the thickness of a surface oxide layer, which is to be expected due to normal oxidation in ambient air and the resist stripping, similarly to the Si$_3$N$_4$ gratings. In addition, the relevant material densities are optimized to scale the optical constants. To take into account uncertainties for the tabulated optical constants\cite{T.Schoonjans2011} and the fluorescence production cross section $\tau_{E_0} \omega_k$, scaling factors with limits between $1 \pm 8\%$ (for $\tau_{E_0} \omega_k$) and $1 \pm 10\%$ for each optical constant are used (for chromium and its oxide). For the calculation, we take into account the incident beam divergence. From this modeling procedure, a height of 25 nm, a width (and length - defined to be identical) of 295 nm and a pitch of 1002 nm is derived. These results agree well with both the nominal values and with the reference-free GIXRF results of randomly ordered sister samples (see ref. \cite{M.Dialameh2017}). For the randomly ordered samples, a very simple reduced-density layer modeling approach was performed that fails for well-ordered structures. The best-fit calculated fluorescence signals, as shown in figs. \ref{fig:Cr_fitcomparison} and \ref{fig:Cr_fitcomparison_Maps}, agree well with the experimental data for various $\varphi$ values.


For the surface oxide layer thickness, the optimization resulted in a value of 11 nm, meaning that almost half of the Cr structure is actually oxidized. Given the fact that the sample was about one year old at the time of the experiment this could be a reasonable result. The small remaining discrepancies visible in figure \ref{fig:Cr_fitcomparison} are likely to be caused by shape deviations from a perfect cuboid structure. Slightly rounded corner imperfections are very likely to be present but have not been taken into account in the modeling procedure so far. 

\begin{figure}[h]
\centering
\includegraphics[width=9cm]{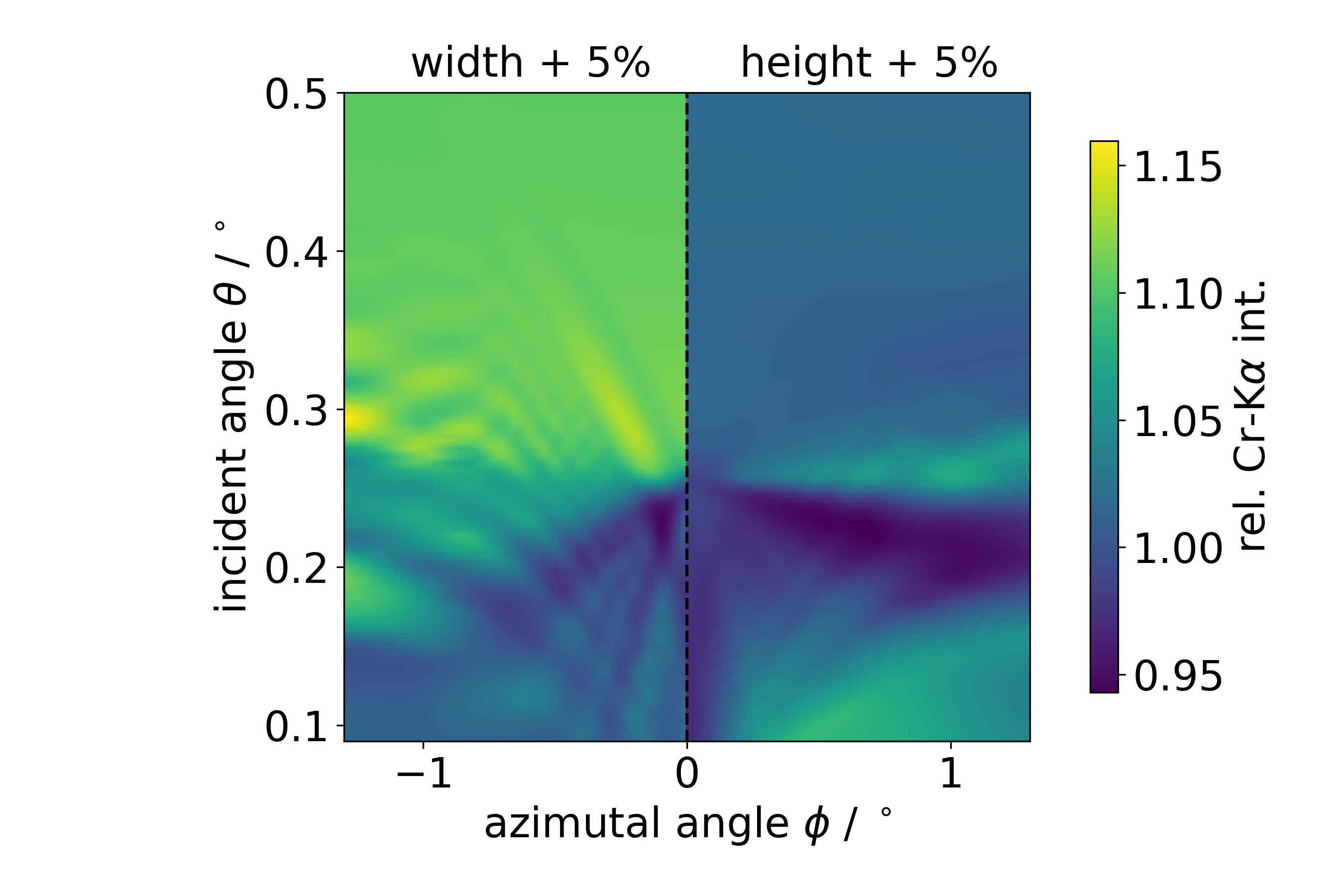}
\caption{Comparison of the fluorescence maps calculated for a 5 \% variation of the nanostructure height and width respectively normalized to the best fit map.}
\label{fig:Cr_Sensitivity_Maps}
\end{figure}

For a demonstration of the sensitivity of the technique, we have performed additional forward simulations with a slightly varied dimension of the nanostructure. If one assumes a height increase by 5 \% (1.2 nm), the detected fluorescence intensity map will be drastically different, especially for high $\varphi$ angles. This is shown in figure \ref{fig:Cr_Sensitivity_Maps} (left side), where the ratio of the fluorescence intensity map with increased height is normalized to the best fit intensity map (as shown in figure \ref{fig:Cr_fitcomparison_Maps}). Similar behaviour can be observed for a 5 \% increase (15 nm) in the nanostructure width, for which the results are shown on the right side of figure \ref{fig:Cr_Sensitivity_Maps}. Due to the automatically larger increase in the total amount of Cr from the increased width, a larger fluorescence signal is observed for high incident angles. In addition, changes are visible for higher $\varphi$ angles. These intensity variations in the fluorescence signal can be easily observed using the technique presented, as they are well above the uncertainties achievable for the experimental data.

\section{Conclusion}
In this work, we have shown that the GIXRF-based characterization of regularly ordered, nanostructured surfaces can be used to derive both dimensional parameters and elemental distributions for 2D and even 3D objects. In contrast to non-ordered objects, more sophisticated approaches are required to model 2D and 3D objects (specifically, to calculate the spatial XSW field distribution). Here, we use either an FEM-based or MB-DDT based calculation scheme in order to model the experimental data. Due to the element sensitivity of the GIXRF technique, angular-dependent fluorescence maps from various elements contained in the structure can be obtained at the same time and a combined modeling of these datasets allows higher discrimination capabilities for the elemental composition of the nanostructure. This is the main benefit of applying GIXRF (as opposed to scattering techniques) when characterizing nanostructures, as it provides non-destructive access to the spatial distribution of the fluorescence-emitting atoms within the structure. Although the examples considered in this work mainly exhibited homogeneous material distributions, a dopant gradient or something comparable would show different angular-dependent fluorescence maps depending on its distribution within the nanostructure; these maps could potentially be characterized using GIXRF. It should also be noted that the underlying modeling techniques can also process more complex nanostructures deviating from box-like shapes such as real FinFET devices or other novel transistor architectures. 

In addition, due to the relatively large irradiated sample area, the GIXRF technique is capable of providing statistically representative data, which is relatively time consuming using comparable techniques for characterization of both dimensional parameters and elemental distributions, e.g. atom probe tomography. These features make GIXRF an interesting nanotechnology metrology tool for the characterization of current and future nanostructures. Furthermore, the technique is in principle also transferable to laboratory-based equipment even though we use synchrotron radiation-based excitation in this work. Several examples (including commercially available tools) of the laboratory-based realization of grazing-incidence XRF instruments already exist; there is no reason why a GIXRF-based characterization of nanostructures should not be transferred to laboratory tools.

\section{Acknowledgements}
P.H., Y.K. and B.B. performed this research within the EMPIR projects Aeromet and Adlab-XMet. The financial support of the EMPIR program is gratefully acknowledged. It is jointly funded by the European Metrology Programme for Innovation and Research (EMPIR) and participating countries within the European Association of National Metrology Institutes (EURAMET) and the European Union.
A.A., V.S. and F.S. acknowledge the funding this project has received from the Electronic Component Systems for European Leadership Joint Undertaking under grant agreement No 826589 — MADEin4. This Joint Undertaking receives support from the European Union's Horizon 2020 research and innovation programme and Netherlands, France, Belgium, Germany, Czech Republic, Austria, Hungary and Israel.





\bibliographystyle{model1-num-names}
\bibliography{sample}

\begin{thebibliography}{33}
\expandafter\ifx\csname natexlab\endcsname\relax\def\natexlab#1{#1}\fi
\providecommand{\bibinfo}[2]{#2}
\ifx\xfnm\relax \def\xfnm[#1]{\unskip,\space#1}\fi
\bibitem[{Pala et~al.(2013)Pala, Liu, Barnard, Askarov, Garnett, Fan, and
  Brongersma}]{Pala2013}
\bibinfo{author}{R.~Pala}, \bibinfo{author}{J.~Liu},
  \bibinfo{author}{E.~Barnard}, \bibinfo{author}{D.~Askarov},
  \bibinfo{author}{E.~Garnett}, \bibinfo{author}{S.~Fan},
  \bibinfo{author}{M.~Brongersma},
\newblock \bibinfo{title}{Optimization of non-periodic plasmonic light-trapping
  layers for thin-film solar cells},
\newblock \bibinfo{journal}{Nat. Commun.} \bibinfo{volume}{4}
  (\bibinfo{year}{2013}) \bibinfo{pages}{2095}.
\bibitem[{Brongersma et~al.(2014)Brongersma, Cui, and Fan}]{M.L.Brongersma2014}
\bibinfo{author}{M.~Brongersma}, \bibinfo{author}{Y.~Cui},
  \bibinfo{author}{S.~Fan},
\newblock \bibinfo{title}{Light management for photovoltaics using high-index
  nanostructures},
\newblock \bibinfo{journal}{Nat Mater} \bibinfo{volume}{13}
  (\bibinfo{year}{2014}) \bibinfo{pages}{451--460}.
\bibitem[{Zheng et~al.(2016)Zheng, Smith, Jackson, Moran, Cui, Chen, Ye, Fang,
  Rodriguez, Weisgraber, and Spadaccini}]{Zheng_2016}
\bibinfo{author}{X.~Zheng}, \bibinfo{author}{W.~Smith},
  \bibinfo{author}{J.~Jackson}, \bibinfo{author}{B.~Moran},
  \bibinfo{author}{H.~Cui}, \bibinfo{author}{D.~Chen}, \bibinfo{author}{J.~Ye},
  \bibinfo{author}{N.~Fang}, \bibinfo{author}{N.~Rodriguez},
  \bibinfo{author}{T.~Weisgraber}, \bibinfo{author}{C.~M. Spadaccini},
\newblock \bibinfo{title}{Multiscale metallic metamaterials},
\newblock \bibinfo{journal}{Nat. Mater.} \bibinfo{volume}{15}
  (\bibinfo{year}{2016}) \bibinfo{pages}{1100--1106}.
\bibitem[{Moore(1965)}]{Moore1965}
\bibinfo{author}{G.~Moore},
\newblock \bibinfo{title}{Cramming more components onto integrated circuits},
\newblock \bibinfo{journal}{Electronics} \bibinfo{volume}{38(8)}
  (\bibinfo{year}{1965}) \bibinfo{pages}{114--117}.
\bibitem[{Wagner and Harned(2010)}]{Wagner_2010}
\bibinfo{author}{C.~Wagner}, \bibinfo{author}{N.~Harned},
\newblock \bibinfo{title}{Lithography gets extreme},
\newblock \bibinfo{journal}{Nature Photonics} \bibinfo{volume}{4}
  (\bibinfo{year}{2010}) \bibinfo{pages}{24--26}.
\bibitem[{King et~al.(2013)King, Simka, Herr, Akinaga, and
  Garner}]{S.W.King2013}
\bibinfo{author}{S.~King}, \bibinfo{author}{H.~Simka},
  \bibinfo{author}{D.~Herr}, \bibinfo{author}{H.~Akinaga},
  \bibinfo{author}{M.~Garner},
\newblock \bibinfo{title}{Research updates: The three m’s (materials,
  metrology, and modeling) together pave the path to future nanoelectronic
  technologies},
\newblock \bibinfo{journal}{APL Materials} \bibinfo{volume}{1}
  (\bibinfo{year}{2013}) \bibinfo{pages}{040701}.
\bibitem[{Natarajan et~al.(2014)Natarajan, Agostinelli, Akbar, Bost, Bowonder,
  Chikarmane, Chouksey, Dasgupta, Fischer, Fu, Ghani, Giles, S.~Govindaraju,
  Han, Hanken, Haralson, Haran, Heckscher, Heussner, Jain, James, Jhaveri, Jin,
  Kam, Karl, Kenyon, Liu, Luo, Mehandru, Morarka, Neiberg, Packan, Paliwal,
  Parker, Patel, Patel, Pelto, Pipes, Plekhanov, Prince, Rajamani, Sandford,
  Sell, Sivakumar, Smith, Song, Tone, Troeger, Wiedemer, Yang, and
  Zhang}]{S.Natarajan2014}
\bibinfo{author}{S.~Natarajan}, \bibinfo{author}{M.~Agostinelli},
  \bibinfo{author}{S.~Akbar}, \bibinfo{author}{M.~Bost},
  \bibinfo{author}{A.~Bowonder}, \bibinfo{author}{V.~Chikarmane},
  \bibinfo{author}{S.~Chouksey}, \bibinfo{author}{A.~Dasgupta},
  \bibinfo{author}{K.~Fischer}, \bibinfo{author}{Q.~Fu},
  \bibinfo{author}{T.~Ghani}, \bibinfo{author}{M.~Giles},
  \bibinfo{author}{R.~G. S.~Govindaraju}, \bibinfo{author}{W.~Han},
  \bibinfo{author}{D.~Hanken}, \bibinfo{author}{E.~Haralson},
  \bibinfo{author}{M.~Haran}, \bibinfo{author}{M.~Heckscher},
  \bibinfo{author}{R.~Heussner}, \bibinfo{author}{P.~Jain},
  \bibinfo{author}{R.~James}, \bibinfo{author}{R.~Jhaveri},
  \bibinfo{author}{I.~Jin}, \bibinfo{author}{H.~Kam},
  \bibinfo{author}{E.~Karl}, \bibinfo{author}{C.~Kenyon},
  \bibinfo{author}{M.~Liu}, \bibinfo{author}{Y.~Luo},
  \bibinfo{author}{R.~Mehandru}, \bibinfo{author}{S.~Morarka},
  \bibinfo{author}{L.~Neiberg}, \bibinfo{author}{P.~Packan},
  \bibinfo{author}{A.~Paliwal}, \bibinfo{author}{C.~Parker},
  \bibinfo{author}{P.~Patel}, \bibinfo{author}{R.~Patel},
  \bibinfo{author}{C.~Pelto}, \bibinfo{author}{L.~Pipes},
  \bibinfo{author}{P.~Plekhanov}, \bibinfo{author}{M.~Prince},
  \bibinfo{author}{S.~Rajamani}, \bibinfo{author}{J.~Sandford},
  \bibinfo{author}{B.~Sell}, \bibinfo{author}{S.~Sivakumar},
  \bibinfo{author}{P.~Smith}, \bibinfo{author}{B.~Song},
  \bibinfo{author}{K.~Tone}, \bibinfo{author}{T.~Troeger},
  \bibinfo{author}{J.~Wiedemer}, \bibinfo{author}{M.~Yang},
  \bibinfo{author}{K.~Zhang},
\newblock \bibinfo{title}{A 14nm logic technology featuring 2nd-generation
  finfet, air-gapped interconnects, self-aligned double patterning and a 0.0588
  $\mu$m$^2$ sram cell size},
\newblock \bibinfo{journal}{Electron Devices Meeting (IEDM), 2014 IEEE
  International}  (\bibinfo{year}{2014}) \bibinfo{pages}{3.7.1 -- 3.7.3}.
\bibitem[{Kim et~al.(2018)Kim, Sherazi, Debacker, Raghavan, Ryckaert, Malik,
  Verkest, Lee, Gillijns, Tan, Blanco, Ronse, and McIntyre}]{Kim_2018}
\bibinfo{author}{R.~R.-H. Kim}, \bibinfo{author}{S.~M.~Y. Sherazi},
  \bibinfo{author}{P.~Debacker}, \bibinfo{author}{P.~Raghavan},
  \bibinfo{author}{J.~Ryckaert}, \bibinfo{author}{A.~Malik},
  \bibinfo{author}{D.~Verkest}, \bibinfo{author}{J.~U. Lee},
  \bibinfo{author}{W.~Gillijns}, \bibinfo{author}{L.~E. Tan},
  \bibinfo{author}{V.~Blanco}, \bibinfo{author}{K.~Ronse},
  \bibinfo{author}{G.~McIntyre},
\newblock \bibinfo{title}{{IMEC} n7, n5 and beyond: {DTCO}, {STCO} and {EUV}
  insertion strategy to maintain affordable scaling trend}
  (\bibinfo{year}{2018}).
\bibitem[{Clark(2014)}]{Clark2014}
\bibinfo{author}{R.~Clark},
\newblock \bibinfo{title}{Emerging applications for high k materials in vlsi
  technology},
\newblock \bibinfo{journal}{Materials} \bibinfo{volume}{7}
  (\bibinfo{year}{2014}) \bibinfo{pages}{2913--2944}.
\bibitem[{Vandooren et~al.(2018)Vandooren, Franco, Wu, Parvais, Li, Witters,
  Walke, Peng, Deshpande, Rassoul, Hellings, Jamieson, Inoue, Devriendt,
  Teugels, Heylen, Vecchio, Zheng, Rosseel, Vanherle, Hikavyy, Mannaert, Chan,
  Ritzenthaler, Mitard, Ragnarsson, Waldron, Heyn, Demuynck, Boemmels, Mocuta,
  Ryckaert, and Collaert}]{Vandooren_2018}
\bibinfo{author}{A.~Vandooren}, \bibinfo{author}{J.~Franco},
  \bibinfo{author}{Z.~Wu}, \bibinfo{author}{B.~Parvais},
  \bibinfo{author}{W.~Li}, \bibinfo{author}{L.~Witters},
  \bibinfo{author}{A.~Walke}, \bibinfo{author}{L.~Peng},
  \bibinfo{author}{V.~Deshpande}, \bibinfo{author}{N.~Rassoul},
  \bibinfo{author}{G.~Hellings}, \bibinfo{author}{G.~Jamieson},
  \bibinfo{author}{F.~Inoue}, \bibinfo{author}{K.~Devriendt},
  \bibinfo{author}{L.~Teugels}, \bibinfo{author}{N.~Heylen},
  \bibinfo{author}{E.~Vecchio}, \bibinfo{author}{T.~Zheng},
  \bibinfo{author}{E.~Rosseel}, \bibinfo{author}{W.~Vanherle},
  \bibinfo{author}{A.~Hikavyy}, \bibinfo{author}{G.~Mannaert},
  \bibinfo{author}{B.~T. Chan}, \bibinfo{author}{R.~Ritzenthaler},
  \bibinfo{author}{J.~Mitard}, \bibinfo{author}{L.~Ragnarsson},
  \bibinfo{author}{N.~Waldron}, \bibinfo{author}{V.~D. Heyn},
  \bibinfo{author}{S.~Demuynck}, \bibinfo{author}{J.~Boemmels},
  \bibinfo{author}{D.~Mocuta}, \bibinfo{author}{J.~Ryckaert},
  \bibinfo{author}{N.~Collaert},
\newblock \bibinfo{title}{First demonstration of 3d stacked finfets at a 45nm
  fin pitch and 110nm gate pitch technology on 300mm wafers},
\newblock \bibinfo{journal}{2018 {IEEE} International Electron Devices Meeting
  ({IEDM})}  (\bibinfo{year}{2018}).
\bibitem[{Ryckaert et~al.(2018)Ryckaert, Schuddinck, Weckx, Bouche, Vincent,
  Smith, Sherazi, Mallik, Mertens, Demuynck, Bao, Veloso, Horiguchi, Mocuta,
  Mocuta, and Boemmels}]{Ryckaert_2018}
\bibinfo{author}{J.~Ryckaert}, \bibinfo{author}{P.~Schuddinck},
  \bibinfo{author}{P.~Weckx}, \bibinfo{author}{G.~Bouche},
  \bibinfo{author}{B.~Vincent}, \bibinfo{author}{J.~Smith},
  \bibinfo{author}{Y.~Sherazi}, \bibinfo{author}{A.~Mallik},
  \bibinfo{author}{H.~Mertens}, \bibinfo{author}{S.~Demuynck},
  \bibinfo{author}{T.~H. Bao}, \bibinfo{author}{A.~Veloso},
  \bibinfo{author}{N.~Horiguchi}, \bibinfo{author}{A.~Mocuta},
  \bibinfo{author}{D.~Mocuta}, \bibinfo{author}{J.~Boemmels},
\newblock \bibinfo{title}{The complementary {FET} ({CFET}) for {CMOS} scaling
  beyond n3},
\newblock \bibinfo{journal}{2018 {IEEE} Symposium on {VLSI} Technology}
  (\bibinfo{year}{2018}).
\bibitem[{Bunday(2016)}]{Bunday2016}
\bibinfo{author}{B.~Bunday},
\newblock \bibinfo{title}{Hvm metrology challenges towards the 5 nm node},
\newblock \bibinfo{journal}{Proc. of SPIE} \bibinfo{volume}{9778}
  (\bibinfo{year}{2016}) \bibinfo{pages}{97780E}.
\bibitem[{Bunday et~al.(2017)Bunday, Solecky, Vaid, Bello, and
  Dai}]{Bunday2017}
\bibinfo{author}{B.~Bunday}, \bibinfo{author}{E.~Solecky},
  \bibinfo{author}{A.~Vaid}, \bibinfo{author}{A.~Bello},
  \bibinfo{author}{X.~Dai},
\newblock \bibinfo{title}{Metrology capabilities and needs for 7nm and 5nm
  logic nodes},
\newblock \bibinfo{journal}{Proc. SPIE 10145}  (\bibinfo{year}{2017})
  \bibinfo{pages}{101450G}.
\bibitem[{Franquet et~al.(2016)Franquet, Douhard, Melkonyan, Favia, Conard, and
  Vandervorst}]{Franquet_2016}
\bibinfo{author}{A.~Franquet}, \bibinfo{author}{B.~Douhard},
  \bibinfo{author}{D.~Melkonyan}, \bibinfo{author}{P.~Favia},
  \bibinfo{author}{T.~Conard}, \bibinfo{author}{W.~Vandervorst},
\newblock \bibinfo{title}{Self focusing {SIMS}: Probing thin film composition
  in very confined volumes},
\newblock \bibinfo{journal}{Applied Surface Science} \bibinfo{volume}{365}
  (\bibinfo{year}{2016}) \bibinfo{pages}{143--152}.
\bibitem[{Vandervorst et~al.(2016)Vandervorst, Fleischmann, Bogdanowicz,
  Franquet, Celano, Paredis, and Budrevich}]{W.Vandervorst2016}
\bibinfo{author}{W.~Vandervorst}, \bibinfo{author}{C.~Fleischmann},
  \bibinfo{author}{J.~Bogdanowicz}, \bibinfo{author}{A.~Franquet},
  \bibinfo{author}{U.~Celano}, \bibinfo{author}{K.~Paredis},
  \bibinfo{author}{A.~Budrevich},
\newblock \bibinfo{title}{Dopant, composition and carrier profiling for 3d
  structures},
\newblock \bibinfo{journal}{Mater. Sci. Semicond. Process.}
  \bibinfo{volume}{62} (\bibinfo{year}{2016}) \bibinfo{pages}{31--48}.
\bibitem[{Sunday et~al.(2015)Sunday, List, Chawla, and Kline}]{Sunday_2015}
\bibinfo{author}{D.~F. Sunday}, \bibinfo{author}{S.~List},
  \bibinfo{author}{J.~S. Chawla}, \bibinfo{author}{R.~J. Kline},
\newblock \bibinfo{title}{Determining the shape and periodicity of
  nanostructures using small-angle x-ray scattering},
\newblock \bibinfo{journal}{Journal of Applied Crystallography}
  \bibinfo{volume}{48} (\bibinfo{year}{2015}) \bibinfo{pages}{1355--1363}.
\bibitem[{Kline et~al.(2017)Kline, Sunday, Windover, and Bunday}]{Kline_2017}
\bibinfo{author}{R.~J. Kline}, \bibinfo{author}{D.~F. Sunday},
  \bibinfo{author}{D.~Windover}, \bibinfo{author}{B.~D. Bunday},
\newblock \bibinfo{title}{X-ray scattering critical dimensional metrology using
  a compact x-ray source for next generation semiconductor devices},
\newblock \bibinfo{journal}{Journal of Micro/Nanolithography, {MEMS}, and
  {MOEMS}} \bibinfo{volume}{16} (\bibinfo{year}{2017}) \bibinfo{pages}{014001}.
\bibitem[{Soltwisch et~al.(2017)Soltwisch, Fern{\'{a}}ndez~Herrero,
  Pfl{\"{u}}ger, Haase, Probst, Laubis, Krumrey, and Scholze}]{Soltwisch2017}
\bibinfo{author}{V.~Soltwisch}, \bibinfo{author}{A.~Fern{\'{a}}ndez~Herrero},
  \bibinfo{author}{M.~Pfl{\"{u}}ger}, \bibinfo{author}{A.~Haase},
  \bibinfo{author}{J.~Probst}, \bibinfo{author}{C.~Laubis},
  \bibinfo{author}{M.~Krumrey}, \bibinfo{author}{F.~Scholze},
\newblock \bibinfo{title}{{Reconstructing detailed line profiles of lamellar
  gratings from GISAXS patterns with a Maxwell solver}},
\newblock \bibinfo{journal}{Journal of Applied Crystallography}
  \bibinfo{volume}{50} (\bibinfo{year}{2017}) \bibinfo{pages}{1524--1532}.
\bibitem[{Dialameh et~al.(2017)Dialameh, Lupi, H\"{o}nicke, Kayser, Beckhoff,
  Weimann, Fleischmann, Vandervorst, Dubcek, Pivac, Perego, Seguini, Leo, and
  Boarino}]{M.Dialameh2017}
\bibinfo{author}{M.~Dialameh}, \bibinfo{author}{F.~F. Lupi},
  \bibinfo{author}{P.~H\"{o}nicke}, \bibinfo{author}{Y.~Kayser},
  \bibinfo{author}{B.~Beckhoff}, \bibinfo{author}{T.~Weimann},
  \bibinfo{author}{C.~Fleischmann}, \bibinfo{author}{W.~Vandervorst},
  \bibinfo{author}{P.~Dubcek}, \bibinfo{author}{B.~Pivac},
  \bibinfo{author}{M.~Perego}, \bibinfo{author}{G.~Seguini},
  \bibinfo{author}{N.~D. Leo}, \bibinfo{author}{L.~Boarino},
\newblock \bibinfo{title}{Development and synchrotron-based characterization of
  al and cr nanostructures as potential calibration samples for 3d analytical
  techniques},
\newblock \bibinfo{journal}{Physica Status Solidi A} \bibinfo{volume}{215}
  (\bibinfo{year}{2017}) \bibinfo{pages}{1700866}.
\bibitem[{Soltwisch et~al.(2018)Soltwisch, H\"{o}nicke, Kayser, Eilbracht,
  Probst, Scholze, and Beckhoff}]{V.Soltwisch2018}
\bibinfo{author}{V.~Soltwisch}, \bibinfo{author}{P.~H\"{o}nicke},
  \bibinfo{author}{Y.~Kayser}, \bibinfo{author}{J.~Eilbracht},
  \bibinfo{author}{J.~Probst}, \bibinfo{author}{F.~Scholze},
  \bibinfo{author}{B.~Beckhoff},
\newblock \bibinfo{title}{Element sensitive reconstruction of nanostructured
  surfaces with finite-elements and grazing incidence soft x-ray fluorescence},
\newblock \bibinfo{journal}{Nanoscale} \bibinfo{volume}{10}
  (\bibinfo{year}{2018}) \bibinfo{pages}{6177--6185}.
\bibitem[{H\"{o}nicke et~al.(2019)H\"{o}nicke, Detlefs, Nolot, Kayser,
  M\"{u}hle, Pollakowski, and Beckhoff}]{Hoenicke2019}
\bibinfo{author}{P.~H\"{o}nicke}, \bibinfo{author}{B.~Detlefs},
  \bibinfo{author}{E.~Nolot}, \bibinfo{author}{Y.~Kayser},
  \bibinfo{author}{U.~M\"{u}hle}, \bibinfo{author}{B.~Pollakowski},
  \bibinfo{author}{B.~Beckhoff},
\newblock \bibinfo{title}{Reference-free grazing incidence x-ray fluorescence
  and reflectometry as a methodology for independent validation of x-ray
  reflectometry on ultrathin layer stacks and a depth-dependent
  characterization},
\newblock \bibinfo{journal}{J. Vac. Sci. Technol., A} \bibinfo{volume}{37}
  (\bibinfo{year}{2019}) \bibinfo{pages}{041502}.
\bibitem[{M\"{u}ller et~al.(2014)M\"{u}ller, H\"{o}nicke, Detlefs, and
  Fleischmann}]{M.Mueller2014}
\bibinfo{author}{M.~M\"{u}ller}, \bibinfo{author}{P.~H\"{o}nicke},
  \bibinfo{author}{B.~Detlefs}, \bibinfo{author}{C.~Fleischmann},
\newblock \bibinfo{title}{Characterization of high-k nanolayers by grazing
  incidence x-ray spectrometry},
\newblock \bibinfo{journal}{Materials} \bibinfo{volume}{7(4)}
  (\bibinfo{year}{2014}) \bibinfo{pages}{3147--3159}.
\bibitem[{Senf et~al.(1998)Senf, Flechsig, Eggenstein, Gudat, Klein, Rabus, and
  Ulm}]{F.Senf1998}
\bibinfo{author}{F.~Senf}, \bibinfo{author}{U.~Flechsig},
  \bibinfo{author}{F.~Eggenstein}, \bibinfo{author}{W.~Gudat},
  \bibinfo{author}{R.~Klein}, \bibinfo{author}{H.~Rabus},
  \bibinfo{author}{G.~Ulm},
\newblock \bibinfo{title}{A plane-grating monochromator beamline for the ptb
  undulators at {BESSY} {II}},
\newblock \bibinfo{journal}{J. Synchrotron Rad.} \bibinfo{volume}{5}
  (\bibinfo{year}{1998}) \bibinfo{pages}{780--782}.
\bibitem[{Beckhoff et~al.(2009)Beckhoff, Gottwald, Klein, Krumrey, M\"{u}ller,
  Richter, Scholze, Thornagel, and Ulm}]{B.Beckhoff2009c}
\bibinfo{author}{B.~Beckhoff}, \bibinfo{author}{A.~Gottwald},
  \bibinfo{author}{R.~Klein}, \bibinfo{author}{M.~Krumrey},
  \bibinfo{author}{R.~M\"{u}ller}, \bibinfo{author}{M.~Richter},
  \bibinfo{author}{F.~Scholze}, \bibinfo{author}{R.~Thornagel},
  \bibinfo{author}{G.~Ulm},
\newblock \bibinfo{title}{A quarter-century of metrology using synchrotron
  radiation by {PTB} in berlin},
\newblock \bibinfo{journal}{Physica Status Solidi (B)} \bibinfo{volume}{246}
  (\bibinfo{year}{2009}) \bibinfo{pages}{1415--1434}.
\bibitem[{Krumrey(1998)}]{Krumrey1998}
\bibinfo{author}{M.~Krumrey},
\newblock \bibinfo{title}{Design of a four-crystal monochromator beamline for
  radiometry at bessy ii},
\newblock \bibinfo{journal}{Journal of Synchrotron Radiation}
  \bibinfo{volume}{5(1)} (\bibinfo{year}{1998}) \bibinfo{pages}{6--9}.
\bibitem[{Beckhoff(2008)}]{Beckhoff2008}
\bibinfo{author}{B.~Beckhoff},
\newblock \bibinfo{title}{Reference-free x-ray spectrometry based on metrology
  using synchrotron radiation},
\newblock \bibinfo{journal}{J. Anal. At. Spectrom.} \bibinfo{volume}{23}
  (\bibinfo{year}{2008}) \bibinfo{pages}{845 -- 853}.
\bibitem[{Lubeck et~al.(2013)Lubeck, Beckhoff, Fliegauf, Holfelder,
  H\"{o}nicke, M\"{u}ller, Pollakowski, Reinhardt, and Weser}]{J.Lubeck2013}
\bibinfo{author}{J.~Lubeck}, \bibinfo{author}{B.~Beckhoff},
  \bibinfo{author}{R.~Fliegauf}, \bibinfo{author}{I.~Holfelder},
  \bibinfo{author}{P.~H\"{o}nicke}, \bibinfo{author}{M.~M\"{u}ller},
  \bibinfo{author}{B.~Pollakowski}, \bibinfo{author}{F.~Reinhardt},
  \bibinfo{author}{J.~Weser},
\newblock \bibinfo{title}{A novel instrument for quantitative nanoanalytics
  involving complementary x-ray methodologies},
\newblock \bibinfo{journal}{Rev. Sci. Instrum.} \bibinfo{volume}{84}
  (\bibinfo{year}{2013}) \bibinfo{pages}{045106}.
\bibitem[{Scholze and Procop(2009)}]{F.Scholze2009}
\bibinfo{author}{F.~Scholze}, \bibinfo{author}{M.~Procop},
\newblock \bibinfo{title}{Modelling the response function of energy dispersive
  x-ray spectrometers with silicon detectors},
\newblock \bibinfo{journal}{X-Ray Spectrom.} \bibinfo{volume}{38(4)}
  (\bibinfo{year}{2009}) \bibinfo{pages}{312--321}.
\bibitem[{Burger et~al.(2008)Burger, Zschiedrich, Pomplun, and
  Schmidt}]{burger_jcmsuite:_2008}
\bibinfo{author}{S.~Burger}, \bibinfo{author}{L.~Zschiedrich},
  \bibinfo{author}{J.~Pomplun}, \bibinfo{author}{F.~Schmidt},
\newblock \bibinfo{title}{{JCMsuite}: {An} {Adaptive} {FEM} {Solver} or
  {Precise} {Simulations} in {Nano}-{Optics}},
\newblock \bibinfo{publisher}{Optical Society of America},
  \bibinfo{year}{2008}, p. \bibinfo{pages}{ITuE4}.
\bibitem[{Schoonjans et~al.(2011)Schoonjans, Brunetti, Golosio, del Rio,
  Sol\'{e}, Ferrero, and Vincze}]{T.Schoonjans2011}
\bibinfo{author}{T.~Schoonjans}, \bibinfo{author}{A.~Brunetti},
  \bibinfo{author}{B.~Golosio}, \bibinfo{author}{M.~S. del Rio},
  \bibinfo{author}{V.~Sol\'{e}}, \bibinfo{author}{C.~Ferrero},
  \bibinfo{author}{L.~Vincze},
\newblock \bibinfo{title}{The xraylib library for x-ray–matter interactions.
  recent developments},
\newblock \bibinfo{journal}{Spectrochim. Acta B} \bibinfo{volume}{66}
  (\bibinfo{year}{2011}) \bibinfo{pages}{776 -- 784}.
\bibitem[{H\"onicke et~al.(2016)H\"onicke, Kolbe, Krumrey, Unterumsberger, and
  Beckhoff}]{P.Hoenicke2016a}
\bibinfo{author}{P.~H\"onicke}, \bibinfo{author}{M.~Kolbe},
  \bibinfo{author}{M.~Krumrey}, \bibinfo{author}{R.~Unterumsberger},
  \bibinfo{author}{B.~Beckhoff},
\newblock \bibinfo{title}{Experimental determination of the oxygen k-shell
  fluorescence yield using thin sio2 and al2o3 foils},
\newblock \bibinfo{journal}{Spectrochim. Acta B} \bibinfo{volume}{124}
  (\bibinfo{year}{2016}) \bibinfo{pages}{94--98}.
\bibitem[{Andrle et~al.(2019)Andrle, Hönicke, Schneider, Kayser,
  Hammerschmidt, Burger, Scholze, Beckhoff, and Soltwisch}]{Andrle_2019}
\bibinfo{author}{A.~Andrle}, \bibinfo{author}{P.~Hönicke},
  \bibinfo{author}{P.-I. Schneider}, \bibinfo{author}{Y.~Kayser},
  \bibinfo{author}{M.~Hammerschmidt}, \bibinfo{author}{S.~Burger},
  \bibinfo{author}{F.~Scholze}, \bibinfo{author}{B.~Beckhoff},
  \bibinfo{author}{V.~Soltwisch},
\newblock \bibinfo{title}{Grazing incidence x-ray fluorescence based
  characterization of nanostructures for element sensitive profile
  reconstruction},
\newblock \bibinfo{journal}{Proc. SPIE - Modeling Aspects in Optical Metrology
  VII} \bibinfo{volume}{11057} (\bibinfo{year}{2019}) \bibinfo{pages}{110570M}.
\bibitem[{Nikolaev et~al.(2020)Nikolaev, Soltwisch, Hönicke, Scholze, de~la
  Rie, Yakunin, Makhotkin, van~de Kruijs, and Bijkerk}]{Nikulaev2019}
\bibinfo{author}{K.~V. Nikolaev}, \bibinfo{author}{V.~Soltwisch},
  \bibinfo{author}{P.~Hönicke}, \bibinfo{author}{F.~Scholze},
  \bibinfo{author}{J.~de~la Rie}, \bibinfo{author}{S.~N. Yakunin},
  \bibinfo{author}{I.~A. Makhotkin}, \bibinfo{author}{R.~W.~E. van~de Kruijs},
  \bibinfo{author}{F.~Bijkerk},
\newblock \bibinfo{title}{A semi-analytical approach for the characterization
  of ordered 3d nano structures using grazing-incidence x-ray fluorescence},
\newblock \bibinfo{journal}{Journal of Synchrotron Radiation}
  \bibinfo{volume}{27} (\bibinfo{year}{2020}) \bibinfo{pages}{386--395}.

\end{thebibliography}







\end{document}